\documentclass[12pt]{article}

\usepackage{epsfig}
\usepackage{amssymb,amsmath}

\newcommand{\bea}{\begin{eqnarray}}
\newcommand{\eea}{\end{eqnarray}}

 \makeatletter
\def\alt{\mathrel{\mathpalette\gl@align<}}
\def\agt{\mathrel{\mathpalette\gl@align>}}
\def\gl@align#1#2{\lower.6ex\vbox{\baselineskip\z@skip\lineskip\z@
\ialign{$\m@th#1\hfil##\hfil$\crcr#2\crcr\sim\crcr}}} \makeatother

\begin{document}
%
\vspace*{1.0cm}

\begin{center}
\baselineskip 20pt {\Large\bf
Higgs Boson Mass Bounds in 
Seesaw Extended Standard Model 
with Non-Minimal Gravitational Coupling 
}
\vspace{1cm}

{\large Bin He$^{a}$\footnote{ E-mail:hebin@udel.edu}, 
Nobuchika Okada$^{b}$\footnote{E-mail:okadan@ua.edu}
and Qaisar Shafi$^{a}$\footnote{ E-mail:shafi@bartol.udel.edu}
} \vspace{.5cm}

{\baselineskip 20pt \it
$^a$Bartol Research Institute, Department of Physics and Astronomy, \\
University of Delaware, Newark, DE 19716, USA \\
\vspace{2mm} 
$^b$Department of Physics and Astronomy, 
University of Alabama, Tuscaloosa, AL 35487, USA 
}
\vspace{.5cm}

\vspace{1.5cm} {\bf Abstract}
\end{center}

In the presence of non-minimal gravitational coupling 
 $ \xi H^\dagger H {\cal R}$ 
 between the standard model (SM) Higgs doublet $H$ 
 and the curvature scalar ${\cal R}$, 
 the effective ultraviolet cutoff scale is given by 
 $\Lambda\approx m_P/\xi$, 
 where $m_P$ is the reduced Planck mass, and 
 $\xi \gtrsim 1$ is a dimensionless coupling constant. 
In type I and type III seesaw extended SM, 
 which can naturally explain the observed solar and 
 atmospheric neutrino oscillations, 
 we investigate the implications of this non-minimal 
 gravitational coupling for the SM Higgs boson mass bounds 
 based on vacuum stability and perturbativity arguments. A lower bound 
 on the Higgs boson mass close to 120 GeV is realized 
 with type III seesaw and $\xi \sim 10-10^3$.
 
\thispagestyle{empty}

\newpage

\addtocounter{page}{-1}

\baselineskip 18pt

The search for the SM Higgs boson is arguably the single most 
 important mission of the LHC. 
According to precision electroweak data 
 and the direct lower mass bound from LEP II, 
 a Higgs boson mass in the range of 
 114.4 GeV $\lesssim m_H \lesssim$ 180 GeV \cite{PDG} is favored. 
If one takes the reduced Planck mass 
 $m_P=2.4 \times 10^{18}$ GeV 
 as a natural cutoff scale of the SM, 
 theoretical considerations  based on vacuum stability 
 and perturbativity arguments narrow the SM Higgs boson mass bounds 
 somewhat, namely 128 GeV $\lesssim m_H \lesssim$ 175 GeV 
 \cite{stability1, Ellis:2009tp}. 
Very recently, it has been reported \cite{Tevatron}
 that the SM Higgs boson mass in the mass range
 158 GeV $\lesssim m_H \lesssim$ 175 GeV is excluded at 95\% C.L. 
 by the direct searches at the Tevatron.

Clearly, if there exists some new physics beyond the SM 
 between the electroweak scale and the reduced Planck scale, 
 it can affect these theoretical Higgs boson mass bounds. 
 The seesaw mechanism is a simple and promising extension 
 of the SM to incorporate the neutrino masses and 
 mixings observed in solar and atmospheric neutrino oscillations. 
There are three main seesaw extensions of the SM, 
 type I \cite{seesawI}, type II \cite{seesawII}, and 
 type III \cite{seesawIII}, in which  new particles, 
 singlet right-handed neutrinos, SU(2) triplet scalar, 
 and SU(2) triplet right-handed neutrinos, respectively, are introduced. 
These new particles contribute to the renormalization group equations 
 (RGEs) at energies higher than the seesaw scale and as a result, 
 the Higgs boson mass bounds can be significantly altered. 
The important implications of seesaw models on the Higgs boson mass bounds have been investigated with the reduced Planck mass cutoff
for the various seesaw models, 
 type I \cite{HMass-typeI, HMass-typeIII}, 
 type II \cite{HMass-typeII} and type III \cite{HMass-typeIII}. 

In general, the non-minimal gravitational coupling 
 between the SM Higgs doublet and the curvature scalar, 
\bea
 \xi H^\dagger H {\cal R} , 
\eea
 can be introduced in the SM. 
This coupling opens up a very intriguing scenario 
 for inflationary cosmology, 
 namely, the possibility that the SM Higgs field may play
 the role of inflation field, and this has been investigated
 in several recent papers 
 \cite{Bezrukov:2008dt}-\cite{Einhorn:2009bh}.
As pointed out in \cite{Naturalness}, 
 in the presence of the non-minimal gravitational coupling, 
 it is natural to identify the effective ultraviolet cutoff scale as 
\bea 
 \Lambda \approx \frac{m_P}{\xi} , 
\eea
 for $\xi \gtrsim 1$, rather than $m_P$. 
Note that the cutoff may depend on the background field value which 
 in our case is of order the electroweak scale (see last refs. in \cite{Bezrukov:2008dt} and \cite{Einhorn:2009bh}).

In this paper, we extend  previous work on 
 the Higgs boson mass bounds in type I and III seesaw extended SM 
 \cite{HMass-typeI, HMass-typeIII} 
 to the case with non-minimal gravitational coupling. 
 The ultraviolet cutoff scale is taken to be 
 $\Lambda=m_P/\xi$ in our analysis. 
We will show that the gravitational coupling as well 
 as type I and III seesaw effects can dramatically alter 
 the vacuum stability and perturbativity bounds 
 on the SM Higgs boson mass. In particular, the vacuum stability
 bound on the Higgs boson mass can be lowered to 120 GeV or so, 
 significantly  below the usual lower bound of about 128 GeV found in   
 the absence of seesaw and with $\xi=0$.

In type I seesaw, three generations of SM-singlet right-handed 
 neutrinos $\psi_i (i=1,2,3)$ are introduced. 
The relevant terms in the Lagrangian are given by 
\bea 
 {\cal L} \supset 
 - y_{ij} \overline{\ell_i} \psi_j H - M_R \overline{\psi^c_i} \psi_i ,   
 \label{Yukawa}  
\eea
 where $\ell_i$ is the $i$-th generation SM lepton doublet. 
For simplicity, we assume in this paper that the three 
 right-handed neutrinos are degenerate in mass ($M_R$). 
At energies below $M_R$, the heavy right-handed neutrinos are integrated out
 and the effective dimension five operator is generated 
 by the seesaw mechanism. 
After electroweak symmetry breaking, the light neutrino mass 
 matrix is obtained as
\bea
  {\bf M}_\nu = 
  \frac{v^2}{2 M_R} {\bf Y}_\nu^T {\bf Y}_\nu ,
\eea
 where $v=246$ GeV is the VEV of the Higgs doublet,
 and ${\bf Y}_\nu = y_{ij}$ is a 3$\times$3 Yukawa matrix.

The basic structure of type III seesaw is similar to type I seesaw, 
 except that instead of the singlet right-handed neutrinos, 
 three generations of fermions which transforms as $({\bf 3}, 0)$ 
 under the electroweak gauge group SU(2)$_L\times$U(1)$_Y$
 are introduced: 
\bea
 \psi_i = \sum_a \frac{\sigma^a}{2} \psi_i^a
=\frac{1}{2}
 \left( \begin{array}{cc}
    \psi^0_i & \sqrt{2} \psi^{+}_i   \\
    \sqrt{2} \psi^{-}_i & - \psi^0_i   \\
 \end{array}\right) .
\eea 
With canonically normalized kinetic terms for the triplet fermions, 
 we replace the SM-singlet right-handed neutrinos of type I seesaw 
 in Eq.~(\ref{Yukawa}) by these SU(2) triplet fermions. 
Assuming degenerate masses ($M_R$) for the three triplet fermions, 
 the light neutrino mass matrix via type III seesaw mechanism 
 is obtained as
\bea
  {\bf M}_\nu = \frac{v^2}{8 M_R} {\bf Y}_\nu^T {\bf Y}_\nu . 
\eea

For a renormalization scale $\mu < M_R$,
 the heavy fermions are decoupled, 
 and there is no effect on the RGEs for the SM couplings. 
However, in the presence of the non-minimal gravitational coupling, 
 a factor $s(\mu)$ defined as 
\bea
 s(\mu) =  
 \frac{1 + \frac{\xi \mu^2}{m_P^2}}
      {1+ (6 \xi + 1 )\frac{\xi \mu^2}{m_P^2}} ,
\eea 
 is assigned to each term in the RGEs associated 
 with the physical Higgs boson loop corrections 
 \cite{Bezrukov:2008dt, Barvinsky:2008ia, Clark:2009dc}. 
In our analysis, we employ 2-loop RGEs for the SM couplings. 
Since the SM beta functions suitably modified with the $s$-factor 
 are known only at 1-loop level, we employ the beta functions 
 with the $s$-factor for 1-loop corrections, 
 while the beta functions for 2-loop corrections are
 without the $s$-factor. 
 We have checked that the effects of the $s$-factor in  
 beta functions for 2-loop corrections are negligible 
 as far as our final results are concerned\cite{SHW}.

For the three SM gauge couplings with 
 a renormalization scale $\mu < M_R$, we have
\bea
 \frac{d g_i}{d \ln \mu} =
 \frac{b_i}{16 \pi^2} g_i^3 +\frac{g_i^3}{(16\pi^2)^2}
  \left( \sum_{j=1}^3 B_{ij}g_j^2 - C_i y_t^2   \right),
\label{gauge}
\eea
 where $g_i$ ($i=1,2,3$) are the SM  gauge couplings,
\bea
b_i = \left(\frac{81+s}{20},-\frac{39-s}{12},-7\right),~~~~
 { B_{ij}} =
 \left(
  \begin{array}{ccc}
  \frac{199}{50}& \frac{27}{10}&\frac{44}{5}\\
 \frac{9}{10} & \frac{35}{6}&12 \\
 \frac{11}{10}&\frac{9}{2}&-26
  \end{array}
 \right),  ~~~~
C_i=\left( \frac{17}{10}, \frac{3}{2}, 2 \right),
\label{beta} 
\eea 
and we have included the contribution 
 from the top Yukawa coupling ($y_t$). 
We use the top quark pole mass $M_t=173.1$ GeV 
 and the strong coupling constant at the Z-pole ($M_Z$) 
 $\alpha_S=0.1193$ \cite{Gfitter}. 
For the top Yukawa coupling, we have 
\bea \label{ty}
 \frac{d y_t}{d \ln \mu}
 = y_t  \left(
 \frac{1}{16 \pi^2} \beta_t^{(1)} + \frac{1}{(16 \pi^2)^2} \beta_t^{(2)}
 \right).
\eea
Here the one-loop contribution is
\bea
 \beta_t^{(1)} =  \left( 4 + \frac{s}{2} \right) y_t^2 -
  \left(
    \frac{17}{20} g_1^2 + \frac{9}{4} g_2^2 + 8 g_3^2
  \right) ,
\label{topYukawa-1}
\eea
while the two-loop contribution is given by \cite{RGE} 
\bea
\beta_t^{(2)} &=&
 -12 y_t^4 +   \left(
    \frac{393}{80} g_1^2 + \frac{225}{16} g_2^2  + 36 g_3^2
   \right)  y_t^2  \nonumber \\
 &&+ \frac{1187}{600} g_1^4 - \frac{9}{20} g_1^2 g_2^2 +
  \frac{19}{15} g_1^2 g_3^2
  - \frac{23}{4}  g_2^4  + 9  g_2^2 g_3^2  - 108 g_3^4 \nonumber \\
 &&+ \frac{3}{2} \lambda^2 - 6 \lambda y_t^2 .
\label{topYukawa-2}
\eea
In solving the RGE for the top Yukawa coupling, 
 its value at $\mu=M_t$ is determined from the relation
 between the pole mass and the running Yukawa coupling
 \cite{Pole-MSbar, Pole-MSbar2},
\bea
  M_t \simeq m_t(M_t)
 \left( 1 + \frac{4}{3} \frac{\alpha_3(M_t)}{\pi}
          + 11  \left( \frac{\alpha_3(M_t)}{\pi} \right)^2
          - \left( \frac{m_t(M_t)}{2 \pi v}  \right)^2
 \right),
\eea
 with $ y_t(M_t) = \sqrt{2} m_t(M_t)/v$, where $v=246$ GeV.
Here, the second and third terms in parenthesis correspond to
 one- and two-loop QCD corrections, respectively,
 while the fourth term comes from the electroweak corrections 
 at one-loop level.

The RGE for the Higgs quartic coupling is given by \cite{RGE},
\bea
\frac{d \lambda}{d \ln \mu}
 =   \frac{1}{16 \pi^2} \beta_\lambda^{(1)}
   + \frac{1}{(16 \pi^2)^2}  \beta_\lambda^{(2)},
\label{self}
\eea
with
\bea
 \beta_\lambda^{(1)} &=& (3 + 9 s^2) \lambda^2 -
 \left(  \frac{9}{5} g_1^2+ 9 g_2^2  \right) \lambda
 + \frac{9}{4}  \left(
 \frac{3}{25} g_1^4 + \frac{2}{5} g_1^2 g_2^2 +g_2^4
 \right) + 12 y_t^2 \lambda  - 12 y_t^4 ,
\label{self-1}
\eea
and
\bea
  \beta_\lambda^{(2)} &=&
 -78 \lambda^3  + 18 \left( \frac{3}{5} g_1^2 + 3 g_2^2 \right) \lambda^2
 - \left( \frac{73}{8} g_2^4  - \frac{117}{20} g_1^2 g_2^2
 - \frac{1887}{200} g_1^4  \right) \lambda - 3 \lambda y_t^4
 \nonumber \\
 &&+ \frac{305}{8} g_2^6 - \frac{289}{40} g_1^2 g_2^4
 - \frac{1677}{200} g_1^4 g_2^2 - \frac{3411}{1000} g_1^6
 - 64 g_3^2 y_t^4 - \frac{16}{5} g_1^2 y_t^4
 - \frac{9}{2} g_2^4 y_t^2
 \nonumber \\
 && + 10 \lambda \left(
  \frac{17}{20} g_1^2 + \frac{9}{4} g_2^2 + 8 g_3^2 \right) y_t^2
 -\frac{3}{5} g_1^2 \left(\frac{57}{10} g_1^2 - 21 g_2^2 \right)
  y_t^2  - 72 \lambda^2 y_t^2  + 60 y_t^6.
\label{self-2}
\eea
The Higgs boson pole mass $m_H$ is determined 
 through one-loop effective potential improved by two-loop RGEs. 
The second derivative of the effective potential 
 at the potential minimum leads to \cite{HiggsPole} 
\bea
  m_H^2 &=& \lambda \zeta^2 v^2 + \frac{3}{64 \pi ^2} \zeta^2 v^2
 \left\lbrace  g^4_2
 \left( \log{\frac{g^2_2 \zeta^2 v^2}{4 \mu^2}} + \frac{2}{3}\right) \right.  
 \nonumber \\
 &+& \frac{1}{2}\left(g^2_2+ \frac{3}{5}g^2_1 \right)^2 
 \left[\log{ \frac{ \left(g^2_2+ \frac{3}{5}g^2_1 \right)
 \zeta^2 v^2 }{4 \mu^2}}+
 \frac{2}{3}\right] - \left. 8y^4_t \log{\frac{y^2_t \zeta^2v^2}{2\mu^2}}
 \right\rbrace ,  
\label{HiggsPole}
\eea
 where $\zeta =\exp{\left( -\int_{M_Z}^\mu 
 \frac{\gamma(\mu)}{\mu} d\mu\right)}$,  
 with the anomalous dimension $\gamma$ of the Higgs doublet 
 evaluated at two-loop level. 
All running parameters are evaluated at $\mu=m_H$, 
 and the Higgs boson mass is determined as the root 
 of this equation. 
We have checked that our results on 
 the Higgs boson mass bounds for the SM case 
 ($\xi=0$ and $M_R \to \infty$) coincide 
 with the ones obtained in recent analysis \cite{Ellis:2009tp}.

For the renormalization scale $\mu \geq M_R$, 
 the SM RGEs should be modified to include contributions
 from the singlet and triplet fermions in type I and III 
 seesaw, respectively, so that the RGE evolution 
 of the Higgs quartic coupling is altered. 
For simplicity, we consider only one-loop corrections 
 from the heavy fermions.

We first consider type I seesaw. 
For $\mu \geq M_R$, the above RGEs are modified as 
\bea 
&& \beta_t^{(1)} \to \beta_t^{(1)}  + {\rm tr}\left[ {\bf S_\nu} \right],  
 \nonumber \\
&& \beta_{\lambda}^{(1)} \to \beta_{\lambda}^{(1)} 
  + 4 {\rm tr}\left[ {\bf S_\nu} \right] \lambda 
  - 4 {\rm tr}\left[ {\bf S_\nu}^2 \right], 
\eea
where ${\bf S_\nu}={\bf Y}_\nu^\dagger {\bf Y}_\nu$, 
 and its corresponding RGE is given by 
\bea
16 \pi^2 \frac{d {\bf S_\nu}}{d \ln \mu} 
 = {\bf S_\nu}
  \left[ 6 y_t^2 + 2 \; {\rm tr}\left[ {\bf S_\nu} \right]
   -\left( \frac{9}{10} g_1^2 +\frac{9}{2} g_2^2 \right)
   + (2 + s) {\bf S_\nu} \right] . 
\eea

We analyze the RGEs numerically and show 
 how the vacuum stability and perturbativity 
 bounds on Higgs boson mass are altered 
 in the presence of type I seesaw and the non-minimal 
 gravitational coupling. 
 As previously noted, because of the gravitational coupling, 
 we set the ultraviolet cutoff as 
 $\Lambda = m_P/\xi$ for $\xi \geq 1$ 
 ($\Lambda = m_P$ as usual if $\xi < 1$). 
We define the vacuum stability bound as the lowest Higgs boson mass
 obtained from the running of the Higgs quartic coupling
 which satisfies the condition $\lambda(\mu) \geq 0$
 for any scale between $m_H \leq \mu \leq \Lambda$.
On the other hand, the perturbativity bound is defined as
 the highest Higgs boson mass obtained from the running
 of the Higgs quartic coupling with the condition
 $\lambda(\mu) \leq 4 \pi$ for any scale
 between $m_H \leq \mu \leq \Lambda$. 

In order to see the effects of the neutrino Yukawa coupling
 on the Higgs boson mass bounds, we first examine
 a toy model with ${\bf Y}_\nu = {\rm diag}(0,0,Y_\nu)$. 
In Figure 1, the vacuum stability and perturbativity 
 bounds on Higgs boson mass as a function of $\xi$ 
 are depicted for various $Y_\nu$ values and 
 a fixed seesaw scale $M_R=10^{13}$ GeV.  
The results for the perturbativity bound are almost insensitive  
 to $Y_\nu$. 
On the other hand, for a fixed $\xi < m_P/M_R$, 
 the vacuum instability bound becomes larger, 
 as $Y_\nu$ is increased. 
For a fixed $Y_\nu$,  
 the vacuum instability bound becomes smaller, 
 as $\xi$ is increased. 
When $\xi > m_P/M_R$ or equivalently $\Lambda < M_R$, 
 the vacuum stability and perturbativity bounds 
 coincides with the SM ones with $\Lambda$, as expected. 
For a fixed cutoff scale $\Lambda >  M_R$, 
 the window for the Higgs boson mass 
 between the vacuum stability and perturbative bounds 
 becomes narrower and is eventually closed 
 as $Y_\nu$ becomes sufficiently large. 
This behavior is shown in Figure 2 for various values of $\xi$. 
Increasing $\xi$ widens  the Higgs mass window for a fixed $Y_\nu$.

It is certainly interesting to consider more realistic cases
 which are compatible with the current neutrino oscillation data.
The light neutrino mass matrix is diagonalized
 by a mixing matrix $U_{MNS}$ such that
\bea
  {\bf M}_\nu = \frac{v^2}{2 M_R} \; {\bf S}_\nu
   = U_{MNS} D_\nu U^T_{MNS},
\label{Mix}
\eea
with $D_\nu ={\rm diag}(m_1, m_2, m_3)$,
 where we have assumed, for simplicity, 
 that the Yukawa matrix ${\bf Y}_\nu$ is real. 
We further assume that the mixing matrix has
 the so-called tri-bimaximal form \cite{hps},
\bea
U_{MNS}=
\left(
\begin{array}{ccc}
\sqrt{\frac{2}{3}} & \sqrt{\frac{1}{3}} & 0 \\
-\sqrt{\frac{1}{6}} & \sqrt{\frac{1}{3}} &  \sqrt{\frac{1}{2}} \\
-\sqrt{\frac{1}{6}} & \sqrt{\frac{1}{3}} & -\sqrt{\frac{1}{2}}
\end{array}
\right) ,
\label{ansatz}
\eea
 which is in very good agreement with the current
 best fit values of the neutrino oscillation data \cite{NuData}.
Let us consider two examples for the light neutrino mass spectrum,
 the hierarchical case and the inverted-hierarchical case.
In the hierarchical case, we have
\bea
 D_\nu \simeq
 {\rm diag}(0,\sqrt{\Delta m_{12}^2}, \sqrt{\Delta m_{23}^2}),
\eea
while for the inverted-hierarchical case, we choose
\bea
 D_\nu \simeq
 {\rm diag}(\sqrt{-\Delta m_{12}^2 + \Delta m_{23}^2},
 \sqrt{\Delta m_{23}^2}, 0).
\eea
 We fix the input values for the solar and atmospheric
 neutrino oscillation data as \cite{NuData}
\bea
 \Delta m_{12}^2 &=& 8.2 \times 10^{-5} \; {\rm eV}^2, \nonumber \\
 \Delta m_{23}^2 &=& 2.4 \times 10^{-3} \; {\rm eV}^2.
 \label{massdiff}
\eea

 From Eqs.~(\ref{Mix})-(\ref{massdiff}), 
 we can obtain the matrix 
\bea 
 {\bf S_\nu}={\bf Y}_\nu^\dagger {\bf Y}_\nu 
 ={\bf Y}_\nu^T {\bf Y}_\nu 
 =\frac{2 M_R}{v^2} U_{MNS} D_\nu U^T_{MNS}.
\eea  

For a given value of $M_R$, we obtain a concrete 
 3$\times$3 matrix at the $M_R$ scale, which is used 
 as an input in the RGE analysis. 
The windows for the Higgs boson pole mass for 
 the hierarchical and inverted-hierarchical cases
 are shown in Figures 3 and 4, respectively. 
As $M_R$ or equivalently the Yukawa couplings become large,
 the window for the Higgs boson mass becomes narrower
 and is eventually closed for a fixed $\xi$. 
In plots for large values of $\xi$, 
 the Higgs boson mass window first narrows, 
 but opens up again, as $M_R$ is increased. 
This is because $M_R$ becomes larger than $\Lambda$ 
 for a sufficiently large $\xi$.

We next consider type III seesaw. 
The analysis is analogous to the type I seesaw case. 
For $\mu \geq M_R$, the RGEs are modified as \cite{HMass-typeIII}
\bea 
&& \beta_t^{(1)} \to \beta_t^{(1)}  
   + \frac{3}{4}{\rm tr}\left[ {\bf S_\nu} \right],  
 \nonumber \\
&& \beta_{\lambda}^{(1)} \to \beta_{\lambda}^{(1)} 
  + 3 {\rm tr}\left[ {\bf S_\nu} \right] \lambda 
  - \frac{5}{4} {\rm tr}\left[ {\bf S_\nu}^2 \right]. 
\eea
The RGE for ${\bf S_\nu}$ is given by 
\bea
16 \pi^2 \frac{d {\bf S_\nu}}{d \ln \mu} 
 = {\bf S_\nu}
  \left[
   6  y_t^2 + \frac{3}{2} {\rm tr}\left[ {\bf S_\nu} \right]
   -\left( \frac{9}{10} g_1^2 +\frac{33}{2} g_2^2 \right)
   + \frac{3+ 2s}{4} \; {\bf S_\nu} \right] . 
\eea
In addition, in type III seesaw, the one-loop beta function 
 coefficient of the SM SU(2) gauge coupling is modified as 
 $-(39-s)/12 \to (9+s)/12$ 
 in the presence of SU(2) triplet fermions.

We first examine the toy model for type III seesaw 
 with $M_R =10^{13}$ GeV. 
The results are depicted in Figure 5, which corresponds 
 to Figure 1 for type I seesaw. 
We can see results similar to those presented in Figure 1. 
The window for the Higgs boson mass 
 between the vacuum stability and perturbativity bounds 
 is shown in Figure 6 for various $\xi$ values, 
 corresponding to Figure 2 for type I seesaw.

In a more realistic case, we repeat the same analysis 
 as in type I seesaw, except for a factor difference 
 in the definition of the light neutrino mass matrix 
 in type III seesaw, 
 ${\bf M}_\nu = \frac{v^2}{8 M_R} \; {\bf S}_\nu$. 
The windows for the Higgs boson pole mass for 
 the hierarchical and inverted-hierarchical cases 
 are shown in Figures 7 and 8, respectively. 
For large $M_R$, 
 we can see behavior similar to Figures 3 and 4 
 for type I seesaw. 
However, note that for low $M_R$ values,
 the Higgs boson mass bounds with type III seesaw 
 are different from the SM ones and the range of 
 the Higgs boson mass window is enlarged, 
 as pointed out in \cite{HMass-typeIII}.
 In particular, a relatively light Higgs boson mass 
 close to 120 GeV is now possible.
This result can be qualitatively understood in the following way.
The presence of the triplet fermions significantly alters 
 the RGE running of the SU(2)$_L$ gauge coupling by making it
 asymptotically non-free, so that $g_2(\mu)$ for $\mu > M_R$
 is larger than the SM value without type III seesaw.
In the analysis of the stability bound, 
 the Higgs quartic coupling is small, 
 and the one-loop beta function of the Higgs quartic coupling 
 can be approximated as (see Eq.~(\ref{self-1})) 
\bea
 \beta_\lambda^{(1)} \simeq 
 \frac{1}{16 \pi^2}
 \left[ \frac{9}{4}  \left(
 \frac{3}{25} g_1^4 + \frac{2}{5} g_1^2 g_2^2 +g_2^4
 \right) - 12 y_t^4 \right].
\eea
Since the first term on the right hand side is larger
 in type III seesaw than in the SM case, 
 the Higgs quartic coupling decreases more slowly than in the SM.
Consequently, the stability bound on the Higgs boson mass is lowered.
For the perturbativity bound, the Higgs quartic coupling is large and
 the one-loop beta function can be approximated by
\bea
 \beta_\lambda^{(1)} \simeq 
  \frac{1}{16 \pi^2} \left[ 
  (3 + 9 s^2) \lambda^2 -
 \left(  \frac{9}{5} g_1^2+9 g_2^2  \right) \lambda
 + 12 y_t^2 \lambda  - 12 y_t^4 \right].
\eea
The beta function is smaller than the SM one due to the second term.
Therefore, the evolution of the Higgs quartic coupling is slower,
 and as a result, the Higgs boson mass based on the perturbative bound
 is somewhat larger than the SM one. 
 
Finally, we note that with type III seesaw, the lower bound on the SM Higgs mass is approximately given by
 \bea
 m_H \geq 121.4 \mathrm{~GeV} 
 +3.0\mathrm{~GeV}\left(\frac{M_t-173.1 \mathrm{~GeV}}{1.3 \mathrm{~GeV}}\right)
 -2.6\mathrm{~GeV}\left(\frac{\alpha _S (M_Z)-0.1193}{0.0028} \right).
 \eea
This is to be compared with a lower bound close to 128 GeV in the absence of 
type III seesaw.

In conclusion, we have considered the potential impacts
 of type I and III seesaw on the vacuum stability 
 and perturbativity bounds on the Higgs boson mass
 in the presence of the non-minimal gravitational coupling, 
 with an effective ultraviolet cutoff scale 
 $\Lambda =m_P/\xi$ for $\xi \geq 1$. 
For energies higher than the seesaw scale, 
 the heavy fermions introduced in type I and III seesaw 
 are involved in loop corrections and 
 the RGEs of the SM are modified. As a consequence, 
 the vacuum stability and perturbativity bounds 
 on the Higgs boson mass are altered.
We have found that for a fixed $\xi$, 
 as the neutrino Yukawa couplings are increased,
 the vacuum stability bound grows and eventually 
 merges with the perturbativity bound. 
Therefore, the Higgs boson mass window is closed 
 at some large Yukawa couplings with a fixed seesaw scale, 
 or some high seesaw scale by fixing the light neutrino mass scale. 
For a fixed neutrino Yukawa coupling 
 or a fixed seesaw scale, the Higgs boson mass window is enlarged  
 as $\xi$ is increased or equivalently 
 the effective cutoff scale is lowered. 
A large neutrino Yukawa coupling or equivalently a large seesaw scale 
 affects in similar ways the Higgs mass bounds in both type I and III seesaw. 
However, with type III seesaw, 
 there is significant lowering of the Higgs mass due to modification
 of the RGE evolution of the SU(2)$_L$ gauge coupling 
 even if the neutrino Yukawa couplings are negligible. 
For a low seesaw scale, the Higgs boson mass window
 between the vacuum stability and perturbative bounds
 turns out to be wider than the SM one.
This is in contrast with type I seesaw
 where the Higgs boson mass bounds in the SM are reproduced
 in the small Yukawa coupling limit.
We have shown that in type III seesaw, 
 the vacuum stability bound on Higgs mass 
 can be close to the current Higgs mass lower bound 
 of 114.4 GeV \cite{LEP2}.

\section*{Acknowledgments}
We thank Ilia Gogoladze and Mansoor Ur Rehman
 for useful comments and discussion. 
This work is supported in part by the DOE Grants, 
 \# DE-FG02-91ER40626 (B.H. and Q.S.) 
 and \# DE-FG02-10ER41714 (N.O.), 
 and by Bartol Research Institute (B.H.). 
N.O. would like to thank the Particle Theory Group 
 of the University of Delaware 
 for hospitality during his visit.


\newpage
\begin{figure}[t]
\includegraphics[scale=1.2]{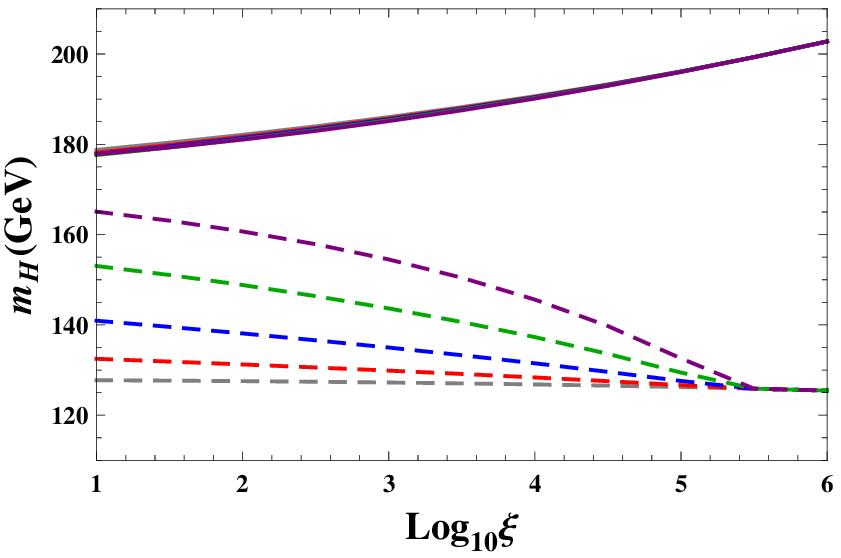}
\caption{
 Perturbativity and vacuum stability bounds 
 on Higgs boson mass versus $\xi$ for various 
 $Y_\nu$ and $M_R=10^{13}$ GeV for type I seesaw. 
 The gray lines correspond to $Y_{\nu}=0$. 
 The red, blue, green and purple lines correspond to 
 $Y_\nu =0.6, 0.8, 1.0$ and $1.2$. 
}
\end{figure}
\begin{figure}[t]
\includegraphics[scale=1.2]{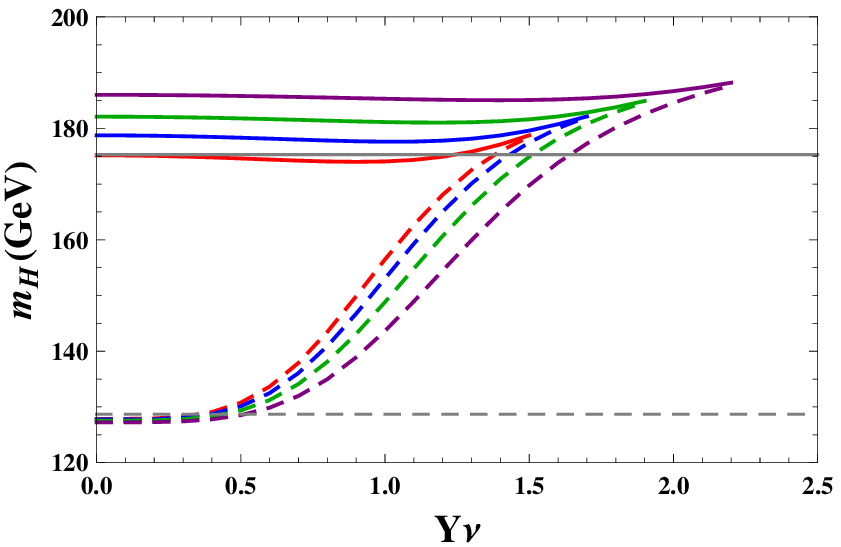}
\caption{
Perturbativity and vacuum stability bounds 
 on Higgs boson mass versus $Y_\nu$ for various 
 $\xi$ and $M_R=10^{13}$ GeV for type I seesaw. 
The red, blue, green and purple lines correspond 
 to $\xi=0,10,100$ and $10^3$. 
The gray lines show the bounds in the SM case.} 
\end{figure}
\begin{figure}[t]
\includegraphics[scale=1.2]{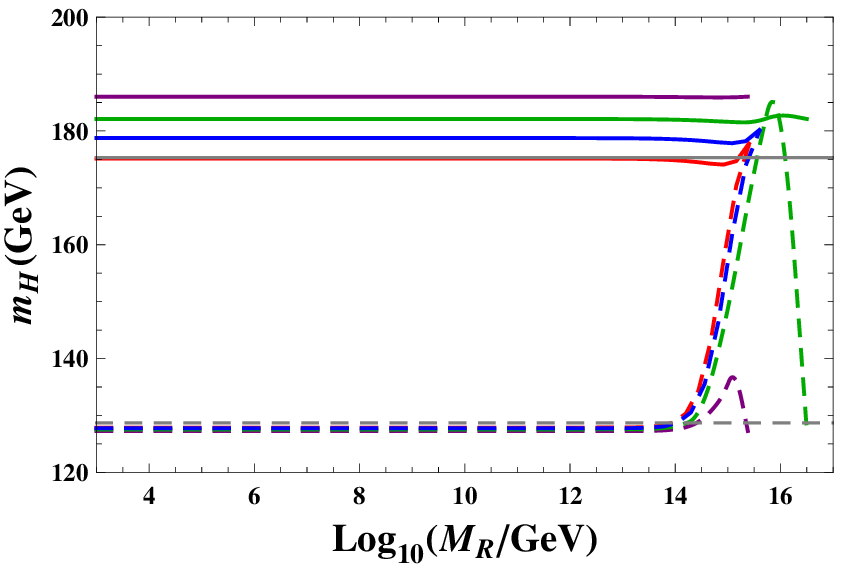}
\caption{ 
Perturbativity and vacuum stability bounds 
 on Higgs boson mass versus $M_R$ 
 with a hierarchical mass spectrum for type I seesaw. 
The red, blue, green and purple lines correspond 
 to $\xi=0,10,100$ and $10^3$. 
The gray lines show the bounds in the SM case.
}
\end{figure}
\begin{figure}[t]
\includegraphics[scale=1.2]{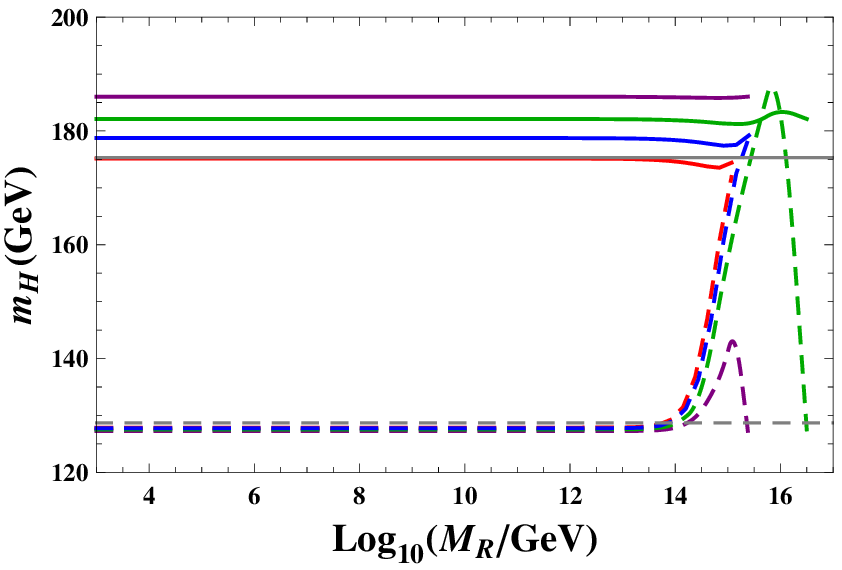}
\caption{ 
Perturbativity and vacuum stability bounds 
 on Higgs boson mass versus $M_R$ 
 with an inverted hierarchical mass spectrum for type I seesaw. 
The red, blue, green and purple lines correspond 
 to $\xi=0,10,100$ and $10^3$. 
The gray lines show the bounds in the SM case.
}
\end{figure}
\begin{figure}[t]
\includegraphics[scale=1.2]{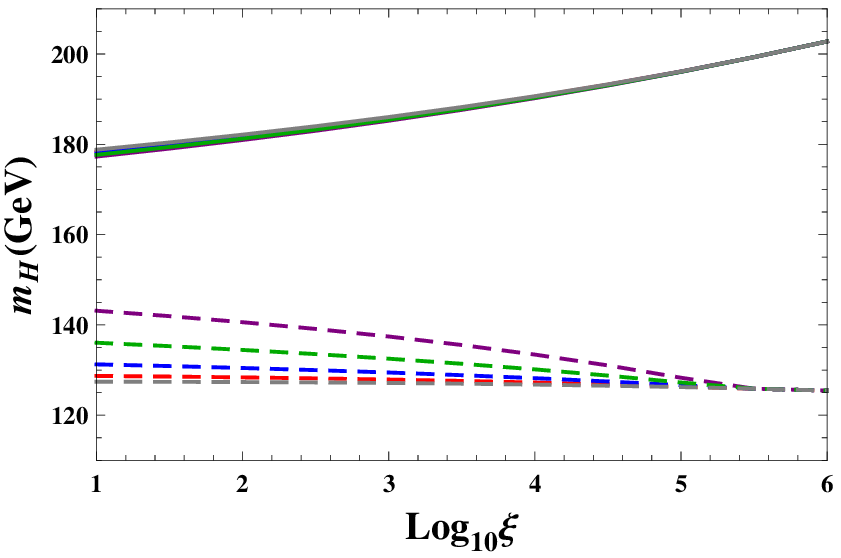}
\caption{
 Perturbativity and vacuum stability bounds 
 on Higgs boson mass versus $\xi$ for various 
 $Y_\nu$ and $M_R=10^{13}$ GeV for type III seesaw. 
 The gray lines correspond to $Y_{\nu}=0$. 
 The red, blue, green and purple lines correspond to 
 $Y_\nu =0.6, 0.8, 1.0$ and $1.2$. 
}
\end{figure}
\begin{figure}[t]
\includegraphics[scale=1.2]{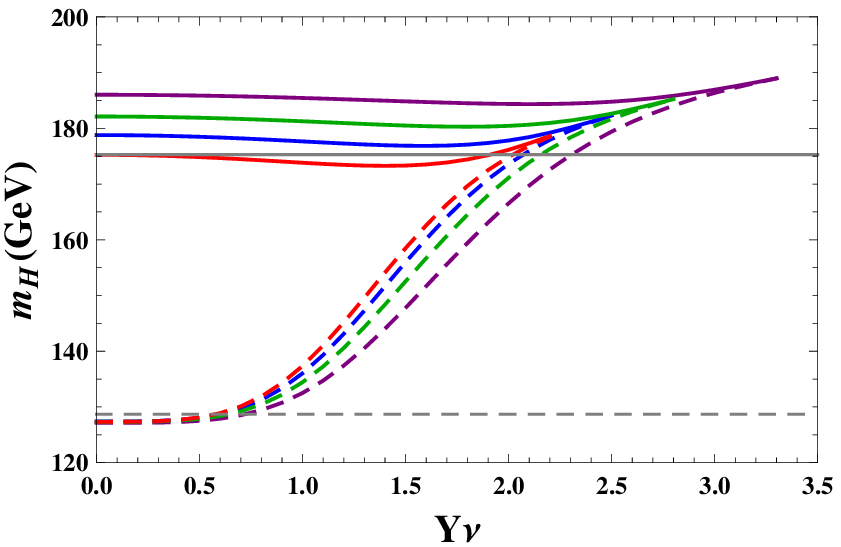}
\caption{
Perturbativity and vacuum stability bounds 
 on Higgs boson mass versus $Y_\nu$ for various 
 $\xi$ and $M_R=10^{13}$ GeV for type III seesaw. 
The red, blue, green and purple lines correspond 
 to $\xi=0,10,100$ and $10^3$. 
The gray lines show the bounds in the SM case.} 
\end{figure}
\begin{figure}[t]
\includegraphics[scale=1.2]{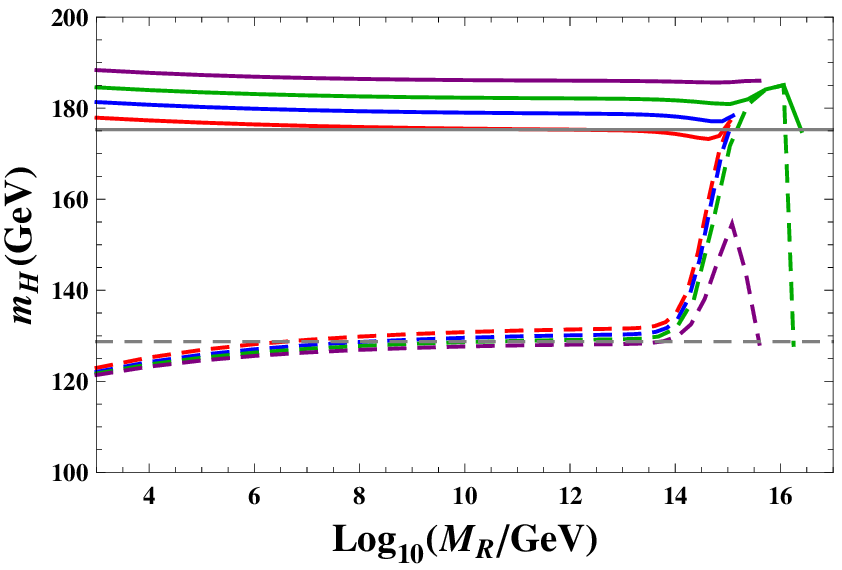}
\caption{ 
Perturbativity and vacuum stability bounds 
 on Higgs boson mass versus $M_R$ 
 with a hierarchical mass spectrum for type III seesaw. 
The red, blue, green and purple lines correspond 
 to $\xi=0,10,100$ and $10^3$. 
The gray lines show the bounds in the SM case.
}
\end{figure}
\begin{figure}[t]
\includegraphics[scale=1.2]{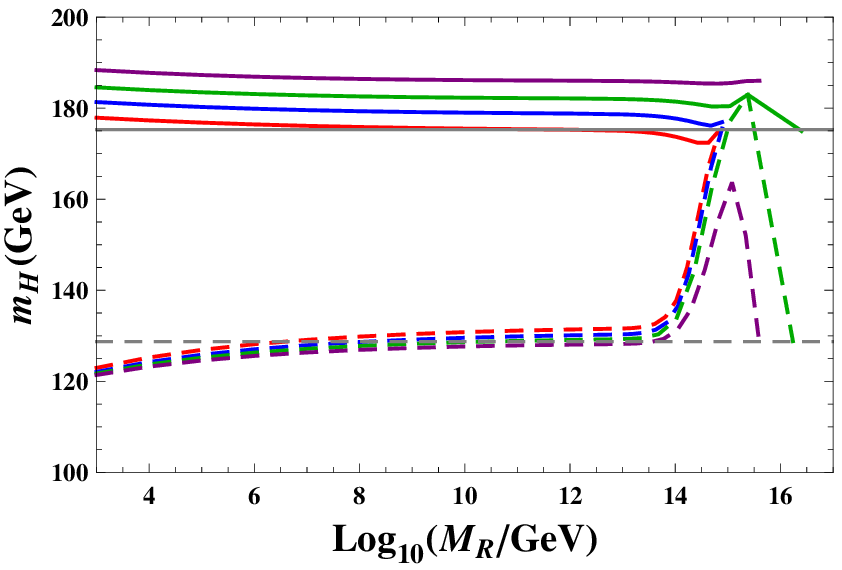}
\caption{ 
Perturbativity and vacuum stability bounds 
 on Higgs boson mass versus $M_R$ 
 with an inverted hierarchical mass spectrum for type III seesaw. 
The red, blue, green and purple lines correspond 
 to $\xi=0,10,100$ and $10^3$. 
The gray lines show the bounds in the SM case.
}
\end{figure}

\end{document}